\documentclass{article}
\usepackage{revsymb}
\usepackage{graphicx}
\usepackage{bm}
\usepackage{amsmath}
\usepackage{amssymb}
\begin{document}

\title{Power spectrum in the Chaplygin gas model:\\
tachyonic, fluid and scalar field representations}
\author{Carlos Eduardo Magalh\~aes Batista \footnote{e-mail: cedumagalhaes@hotmail.com}, J\'{u}lio C.
Fabris\footnote{e-mail: fabris@pq.cnpq.br}\\
Departamento de F\'{\i}sica, \\
Universidade Federal do Esp\'{\i}rito Santo, \\
CEP 29060-900 Vit\'{o}ria, Esp\'{\i}rito Santo, Brasil\\
and\\
Masaaki Morita\footnote{e-mail: morita@mst.nagaokaut.ac.jp}\\
Department of Material Science and Technology,\\
Nagaoka University of Technology,\\
1603-1 Kamitomioka, Nagaoka, Niigata 940-2188, Japan
}
\date{}
\maketitle

\begin{abstract}
The Chaplygin gas model, characterized by an equation of state of the type
$p = - \frac{A}{\rho}$ emerges naturally from the Nambu-Goto action of string
theory. This fluid representation can be recast under the form of a tachyonic
field given by a Born-Infeld type Lagrangian. At the same time,
the Chaplygin gas equation of state can be obtained from a self-interacting scalar field.
We show that, from the point of view of the supernova type Ia data, the three representations (fluid, tachyonic, scalar field) lead to the same results. However, concerning the matter
power spectra, while the fluid and tachyonic descriptions lead to exactly the same
results, the self-interacting scalar field representation implies different statistical estimations for the parameters. In particular, the estimation for the dark matter density parameter in the fluid representation favors a
universe dominated almost completely by dark matter, while in the self-interacting scalar field representation the prediction
is very closed to that obtained in the
$\Lambda$CDM model.

\end{abstract}

\vspace{0.5cm} \leftline{PACS: 98.80.-k, 04.62.+v}

\section{Introduction}

The Claplygin gas equation of state, given by 
\begin{equation}
\label{eos}
 p = - \frac{A}{\rho},
\end{equation}
was first presented in the reference \cite{chaplygin} in the context of the study of 
incompressible fluids. Later, it has revealed important in the study of the aerodynamics
problem \cite{aero}. More recently, it has re-appeared as an example of a irrotational fluid that
can be supersymmetrized \cite{super}. Moreover, it has been discovered an interesting connection between
this equation of state and the Nambu-Goto action of string theory written in the light-cone
variables \cite{jackiw}. Its application to cosmology was first pointed out in \cite{moschella}. A phenomenological generalization has been proposed
\cite{bertolami}, such that the equation of
state takes the form
\begin{equation}
 p = - \frac{A}{\rho^\alpha},
\end{equation}
where $\alpha$ is a free parameter, generally a positive number.
\par
The interest for the Chaplygin gas model, and its phenomenological generalization, in cosmology emerged from the evidences in favor of an accelerating expansion of
the universe at
present time \cite{s1,s2}. These evidences ask for a fluid with negative pressure that begins to dominate the matter content of
the universe at about $z \sim 1$ ($z$ being the cosmological redshift), whithout affecting
the previous history of the universe. In particular the primordial nucleosynthesis, the
transition from a radiation dominated universe to a matter dominated universe and,
finally, the process of structure formation driven by cold dark matter must remain unaltered. In spite of
being a natural candidate for the fluid responsible for the cosmic acceleration, the cosmological constant faces many theoretical and observational problems, mainly represented
by the huge discrepancy of the observed and the predicted values, as well as the necessary fine
tunning of its value to achieve an acceleration just as measured at present time.
\par
The Chaplygin gas
has revealed an interesting alternative to the so-called $\Lambda$CDM model.
In fact, considering a Friedmann-Lema\^{\i}tre-Robertson-Walker universe (FLRW), a fluid
with energy density $\rho$ and pressure $p$ must satisfy the conservation law
\begin{equation}
 \dot\rho + 3\frac{\dot a}{a}(\rho + p) = 0,
\end{equation}
where $a$ is the scale of factor of the spatial section. For the equation of state (\ref{eos}), this equation can be easily integrated leading to
\begin{equation}
\label{sol1}
 \rho = \sqrt{A + \frac{B}{a^6}},
\end{equation}
where $B$ is an integration constant.
The important point concerning (\ref{sol1}) is the fact that it contains relevant
asymptotic regimes: as $a \rightarrow 0$, i.e. as we go back to the past, $\rho \sim a^{-3}$,
reproducing a pressureless fluid (cold dark matter); moreover, as $a \rightarrow \infty$,
$\rho \sim$ constant, leading to a behaviour typical of a cosmological constant.
Hence, the transition from a non-accelerating to an accelerating universe,
can be achieved dynamically. At same time, the process of structure formation is
guaranteeded by the fact that the Chaplygin gas mimics cold dark matter at previous time.
\par
The confrontation of the Chaplygin gas model with observations have shown that
it is quite competitive with other dark matter/dark energy models, like the
$\Lambda$CDM and quintessence models. However, there are some interesting aspects
that point out to the necessity of more deep investigations. For example, while the
confrontation with the supernova type Ia (SN Ia) data shows that the unification model is
preferred \cite{sn1}, that is essentially no exotic new component besides the Chaplygin gas, the structure formation analysis reveal that a large amount of extra
dark matter is still needed \cite{piattella,hermano}. Moreover, supernova data seems to favor, in the case
of the generalized Chaplygin gas model, negatives values of the parameter $\alpha$. If we
turn now structure formation, the study of the cases with negative values of $\alpha$
becomes more sensible since this implies imaginary sound velociy, hence plagued
with instabilities \cite{martin}. Finally, mainly in the case of structure formation, the results
may be sensitive on how the Chaplygin gas is represented, that is, if it is viewed as
a tachyonic field, a fluid or a self-interacting scalar field.
\par
Our goal here is to consider this last problem: how the use of different representations
of the Chaplygin gas affects the final estimation of the cosmological parameters. We use
the SN Ia data and the 2dFGRS data for the matter power spectrum. In special,
we will show that the tachyonic and fluid representations give the same results even
for the matter power spectrum. However, the self-interacting scalar field, even if it admits
in principle an extension for negative values for the parameter $\alpha$ when the generalized Chaplygin gas
is considered, leads
to very different parameter estimations.
\par
From now on, for the sake of simplicity, we will work only with the traditional Chaplygin
gas model defined by the equation of state (\ref{eos}). The tachyonic formulation can
be obtained from the Lagrangian
\begin{equation}
\label{lag-tach}
 L = \sqrt{-g}V(T)\sqrt{1 - T_{;\rho}T^{;\rho}},
\end{equation}
where $T$ is the tachyonic field. The corresponding energy-momentum tensor is then given
by
\begin{equation}
\label{tachyon}
 T_{\mu\nu} = \frac{V(T)}{\sqrt{1 - T_{;\rho}T^{;\rho}}}T_{;\mu}T_{;\nu} + 
V(T)\sqrt{1 - T_{;\rho}T^{;\rho}}g_{\mu\nu}.
\end{equation}
For the case of the FLRW fllat universe, this energy-moment tensor leads to the following
expressions for the energy density and for the pressure:
\begin{eqnarray}
 \rho_T &=& \frac{V(T)}{\sqrt{1 - \dot T^2}};\\
p_T &=& - V(T)\sqrt{1 - \dot T^2} \quad .
\end{eqnarray}
Thus, equation (\ref{eos}) with $A =$ constant implies that $V(T)$ must be a constant, that is,
independent of $T$. For some interesting generalizations with a non-constant $V(T)$ see, for example, \cite{moschellabis}.
\par
Let us consider now the energy-momentum tensor of a perfect fluid, 
\begin{equation}
 T_{\mu\nu} = (\rho + p)u_\mu u_\nu - pg_{\mu\nu},
\end{equation}
where $u_\mu$ is the four-velocity of the fluid. Introducing small perturbations around
a background value ($\delta\rho$, $\delta p$, $\delta u^\mu$ and $\delta g_{\mu\nu} = h_{\mu\nu}$), using the synchronous coordinate condition $h_{\mu0} = 0$, we obtain at first order the following expressions:
\begin{equation}
 \delta T_{00} = \delta\rho, \quad \delta T_{ij} = - p h_{ij} - \delta p g_{ij}.
\end{equation}
Using now the expression (\ref{tachyon}), we obtain
\begin{eqnarray}
 \delta T_{00} &=& - \frac{V \dot T \delta \dot T}{(1 - \dot T^2)^{3/2}} = \delta \rho_T, \\
\delta T_{ij} &=& - \frac{V \dot T\delta\dot T}{\sqrt{1 - \dot T^2}}g_{ij} +
V\sqrt{1 - \dot T^2}h_{ij}= - p_T h_{ij}
- \delta p_T g_{ij}.
\end{eqnarray}
Hence, the fluid approach represented by (\ref{eos}) reproduces the features of the 
tachyonic model described by (\ref{lag-tach})
at the background and at the perturbative levels.
\par
Since the tachyonic fluid is entirely acquainted by the fluid represntation (at classical
level, of course), let us consider the Einstein's equation coupled to baryonic matter and
the Chaplygin gas.
The field equations read,
\begin{eqnarray}
\label{fe1}
 R_{\mu\nu} = 8\pi G\biggr(T^b_{\mu\nu} - \frac{1}{2}g_{\mu\nu}T^b\biggl)
&+& 8\pi G\biggr(T^c_{\mu\nu} - \frac{1}{2}g_{\mu\nu}T^c\biggl),\\
\label{fe2}
{T_c^{\mu\nu}}_{;\mu} = 0, \quad & &{T_b^{\mu\nu}}_{;\mu} = 0 .
\end{eqnarray}
The superscript (subscript) $b$ and $c$ stands for {\it baryons} and {\it Chaplygin}.
\par 
We perturb now these field equations, using again the synchronous coordinate conditon.
We divide all equations by the Hubble parameter today $H_0$ in order to make the equations dimensionless. We end up with the following set of coupled equations for the perturbed quantites:
\begin{eqnarray}
 \ddot\delta_m + 2\frac{\dot a}{a}\dot\delta_m - \frac{3}{2}\Omega_m\delta_m &=& \frac{3}{2}
\Omega_c\frac{4\bar A a^6 + 1 - \bar A}{\bar Aa^6 + (1 - \bar A)}\delta_c, \\
\dot\delta_c + \frac{1 - \bar A}{\bar A a^6 + 1 - \bar A}\theta_c +
6\frac{\dot a}{a}\frac{\bar A a^6}{\bar A a^6 + 1 - \bar A}\delta_c &=& \frac{1 - \bar A}{\bar A a^6 + 1 - \bar A}\dot\delta_m,\\
\dot\theta_c + \frac{\dot a}{a}\frac{\bar A a^6 + 2(1 - \bar A)}{\bar A a^6 + 1 - \bar A}\theta_c &=& \frac{\bar A a^4}{1 - \bar A}k^2\delta_c.
\end{eqnarray}
In these expressions $\delta_m$ and $\delta_c$ are the density contrast for
baryons and for the Chaplygin gas respectively, while $\theta_c$ is the perturbation in the
four velocity connected with the Chaplygin gas. We have noted $\bar A = \frac{A}{\rho_{c0}^2}$, which is the sound velocity of the Chaplygin gas fluid,
with $\rho_{c0} = \sqrt{A + B}$ being the corresponding energy density today.
\par
Passing now from the (dimensionless) time variable $t$ to the scale factor variable $a$, the above equations take the following form:
\begin{eqnarray}
\delta_m'' + \biggr(\frac{2}{a} + \frac{g}{f^2}\biggl)\delta_m' - \frac{3}{2}\frac{\Omega_m}{f^2}\delta_m &=& \frac{3}{2}
\frac{\Omega_c}{f^2}\frac{4\bar A a^6 + 1 - \bar A}{\bar Aa^6 + (1 - \bar A)}\delta_c, \\
\delta_c' + \frac{1 - \bar A}{\bar A a^6 + 1 - \bar A}\frac{\theta_c}{f} +
6\frac{\bar A a^5}{\bar A a^6 + 1 - \bar A}\delta_c &=& \frac{1 - \bar A}{\bar A a^6 + 1 - \bar A}\delta_m',\\
\label{pe3a}
\theta_c' + \frac{\bar A a^6 + 2(1 - \bar A)}{\bar A a^6 + 1 - \bar A}\frac{\theta_c}{a} &=& \frac{\bar A a^4}{1 - \bar A}\frac{k^2l_0^2}{f}\delta_c.
\end{eqnarray}
The primes mean derivative with respect to $a$ and $l_0$ is the Hubble radius today. Moreover, we have the following definitions:
\begin{eqnarray}
f(a) &=& \sqrt{\Omega_m(a) + \Omega_c(a)}a,\\
g(a) &=& - \frac{\Omega_m(a)}{2}a - \frac{\Omega_c(a)}{2}\frac{1 - \bar A - 2\bar Aa^6}{\bar A a^6 + 1 - \bar A}a, \\
\Omega_m &=& \frac{\Omega_0}{a^3},\\
\Omega_c &=& \Omega_{c0}\sqrt{\bar A + \frac{1 - \bar A}{a^6}}.
\end{eqnarray}
Note that the sound speed of the fluid plays an important r\^ole, in particular multiplying the term originated from the gradient of pressure in equation (\ref{pe3a}). It is this fact that leads to instabilities problems when the generalized Chaplygin gas model, with $\alpha$ negative, is
considered.
\begin{center}
\begin{figure}[!t]
\begin{minipage}[t]{0.3\linewidth}
\includegraphics[width=\linewidth]{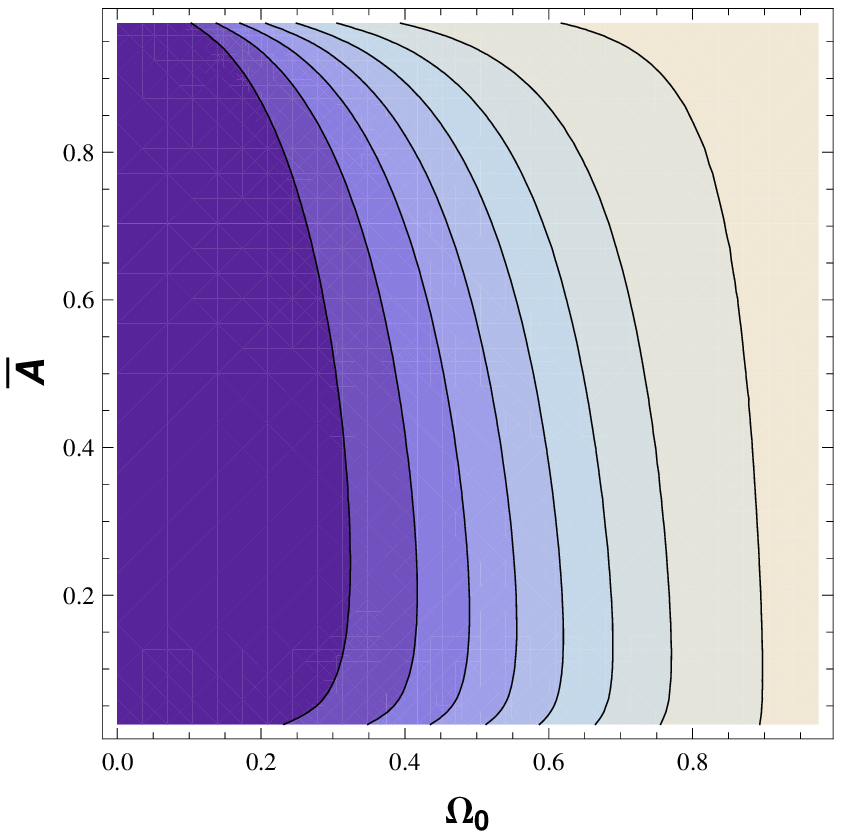}
\end{minipage} \hfill
\begin{minipage}[t]{0.3\linewidth}
\includegraphics[width=\linewidth]{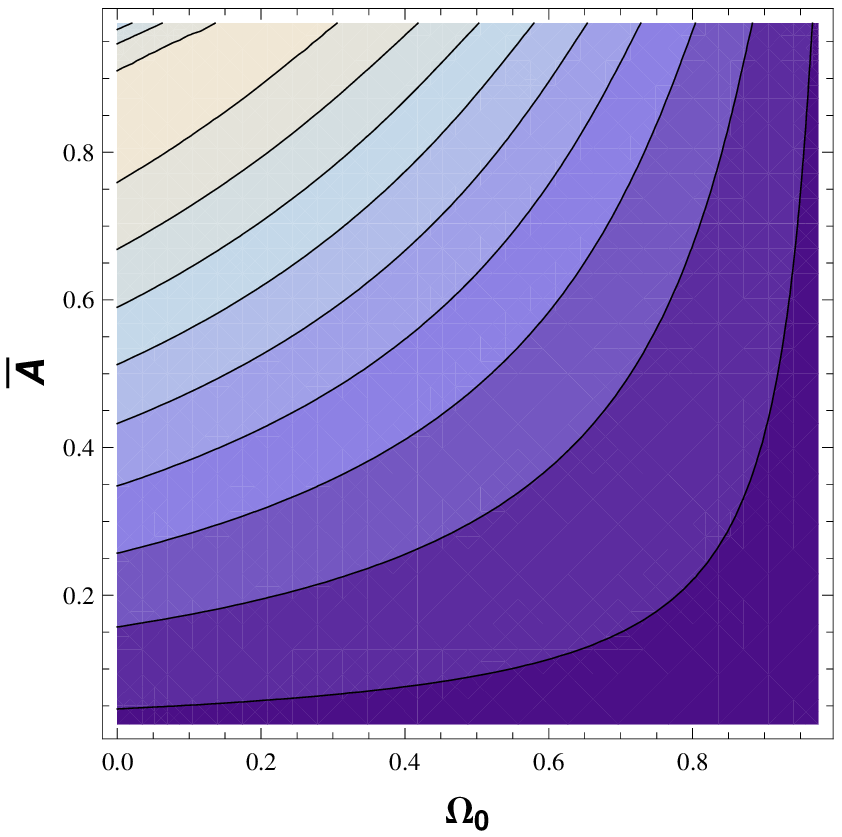}
\end{minipage} \hfill
\begin{minipage}[t]{0.3\linewidth}
\includegraphics[width=\linewidth]{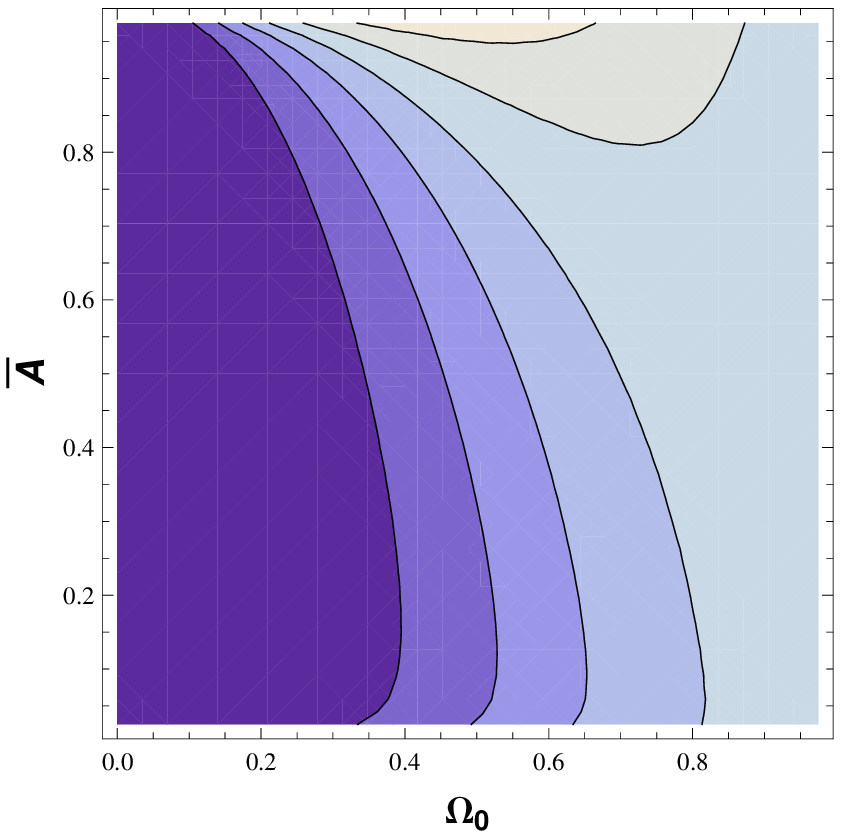}
\end{minipage} \hfill
\caption{{\protect\footnotesize Two-dimensional probability distribution for $\Omega_{0}$ and $\bar A$ using only
the 2dFGRS data (left), only the SN Ia data (center) and the joint probability from both sets of data. The lighter is the region, the higher is
the associated probability density.}}
\end{figure}
\end{center}
\par
The analysis for the SN Ia data is made through the computation of the luminosity distance, given by
\begin{equation}
D_L = \frac{c}{H_0}(1 + z)\int_0^z\frac{dz'}{\sqrt{\Omega_m(z') + \Omega_c(z')}},
\end{equation}
where $z$ is the redshift given by $z = - 1 + \frac{1}{a}$.
\par
We note that, concerning the matter power spectrum we have two free parameters: $\Omega_{m0}$ (or equivalently $\Omega_{c0}$, since
$\Omega_{m0} + \Omega_{c0} = 1$) and $\bar A$. For the SN Ia analysis we have one more free parameter: the Hubble reduced parameter $h$ defined
by $H_0 = 100\,h\,km/Mpc.s$. We use the $\chi^2$ statistical function which gives the quality of the fitting of the observational data
by the theoretical model (see \cite{sn1}). Marginalizing the probability distribution function for the SN Ia data in the variable $h$ (it means,
integrating in this variable),
we end up with two sets of observational constraints, for the 2dFGRS and SN Ia data, depending on the parameters $\bar A$ and $\Omega_{m0}$. 
\par
We compute now the probability distribution function using the 2dFGRS power spectrum data. The initial conditions are fixed using
the prescription described in reference \cite{sola}. The luminosity distance is also computed, for the gold sample, by using the
method sketched in \cite{sn1}. The results are shown in figure 1, where the two-dimensional PDF using only power spectrum data, only
the SN Ia data and the joint probability are displayed. In figure 2, the corresponding estimations for the one-dimensional PDF for
the matter density today are shown, while in the figure 3 the one-dimensional PDF for the sound velocity $\bar A$ for the three sets
of data is displayed.
The results for the SN Ia data agree with those described in reference \cite{sn1}: the unified scenario, with essentially no dark matter,
is preferred; at the same time, very large values for $\bar A$, that is, the "quasi"-$\Lambda$CDM models, are also preferred.
Using the 2dFGRS data, the conclusions of references \cite{piattella,hermano} are re-obtained: the anti-unified model, with essentially all
matter in the form of dark matter, is preferred; for the parameter $\bar A$, again large values are preferred, but there is also a high probability
distribution near zero. The combination of both data privilegize the scenarios indicated by the 2dFGRS even if some new effects appear like the
maximum value for the PDF of $\Omega_0$ for the joint SN Ia and 2dFGRS data in figure 2. But there is some subtle points concerning
this: if the joint higher dimensional PDF in the parameter space is first constructed, as we have done in the present work, than the results
are those displyaed in the figure $3$; however, if the corresponding
one-dimensional PDF's are combined after the marginalization on the other variables, the conclusions are very different, in the sense that
the SN Ia data are favored. But, we think the first procedure is more consistent.
\begin{center}
\begin{figure}[!t]
\begin{minipage}[t]{0.3\linewidth}
\includegraphics[width=\linewidth]{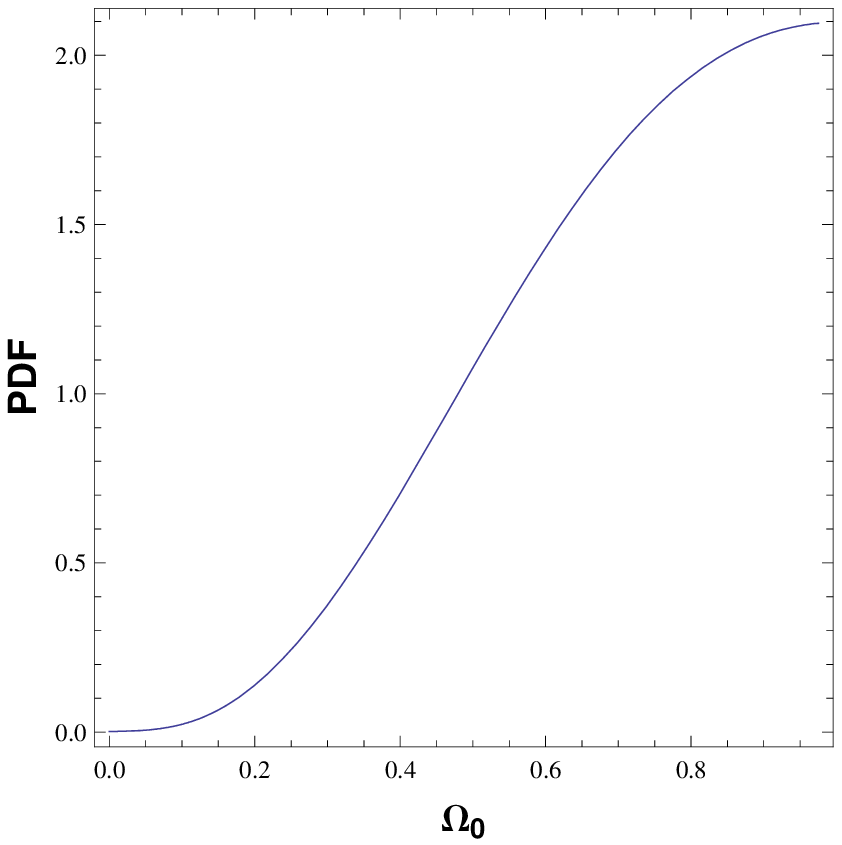}
\end{minipage} \hfill
\begin{minipage}[t]{0.3\linewidth}
\includegraphics[width=\linewidth]{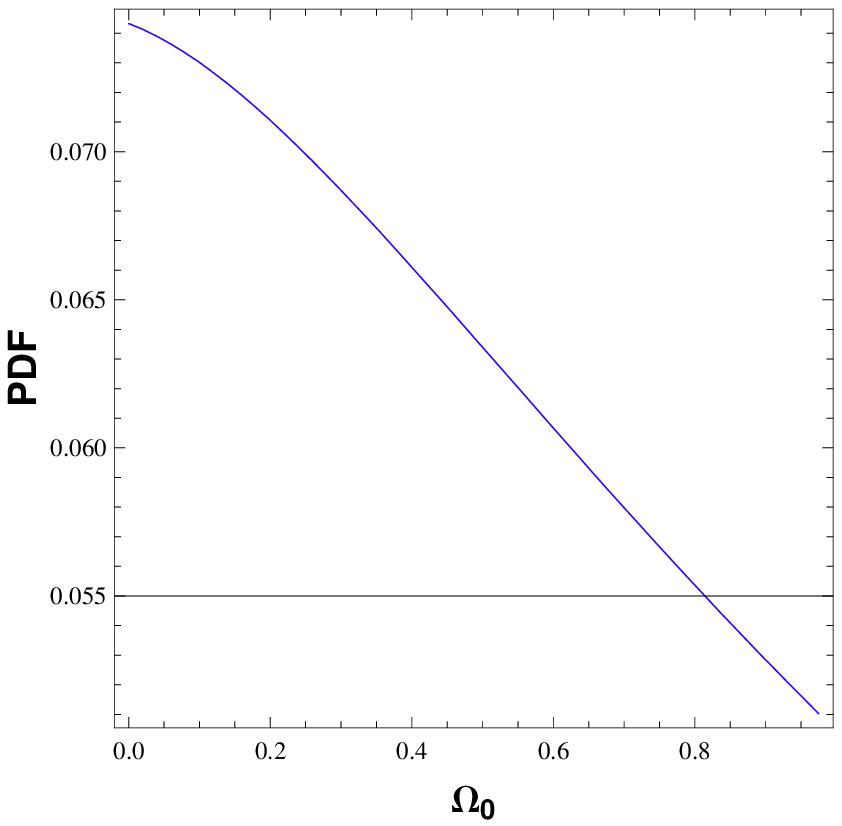}
\end{minipage} \hfill
\begin{minipage}[t]{0.3\linewidth}
\includegraphics[width=\linewidth]{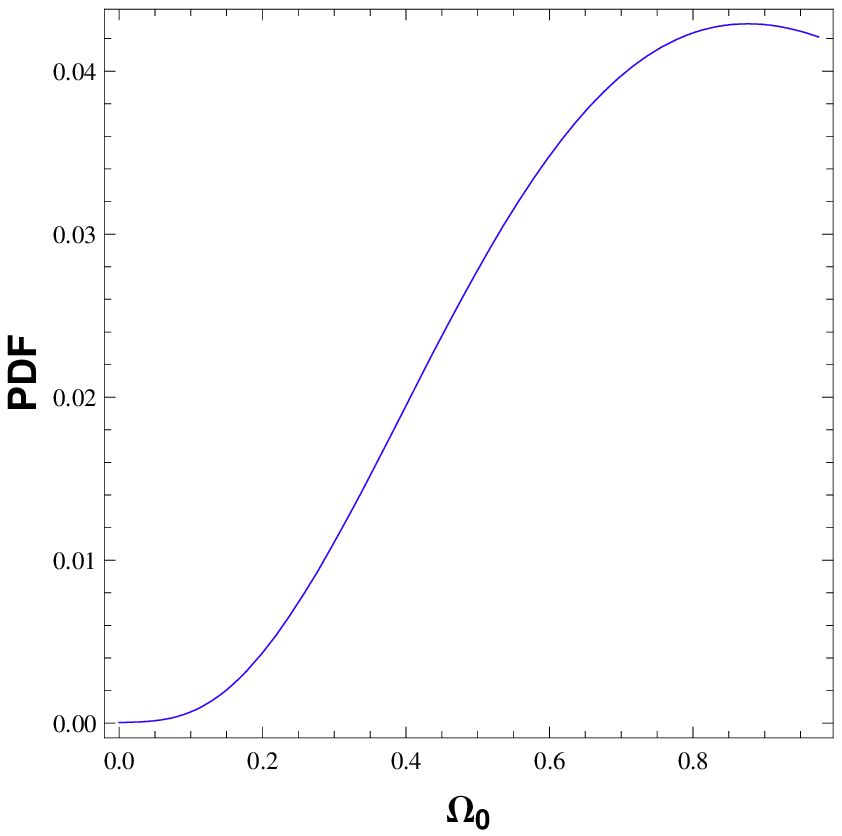}
\end{minipage} \hfill
\caption{{\protect\footnotesize One-dimensional probability distribution for $\Omega_{0}$ using only
the 2dFGRS data (left), only the SN Ia data (center) and the joint probability from both sets of data.}}
\end{figure}
\end{center}\begin{center}
\begin{figure}[!t]
\begin{minipage}[t]{0.3\linewidth}
\includegraphics[width=\linewidth]{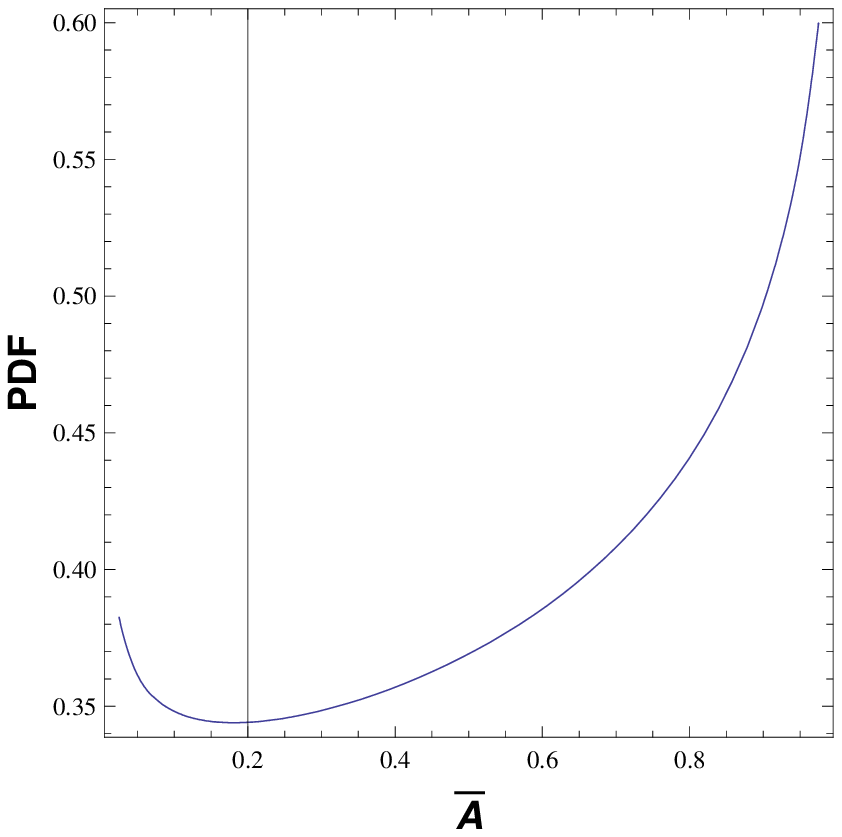}
\end{minipage} \hfill
\begin{minipage}[t]{0.3\linewidth}
\includegraphics[width=\linewidth]{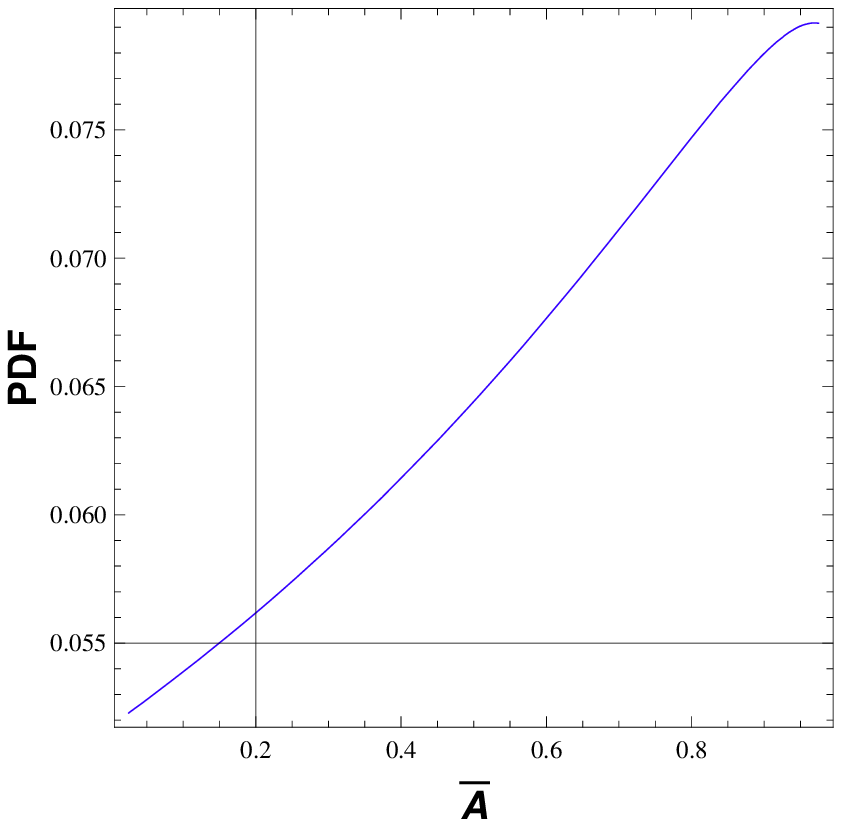}
\end{minipage} \hfill
\begin{minipage}[t]{0.3\linewidth}
\includegraphics[width=\linewidth]{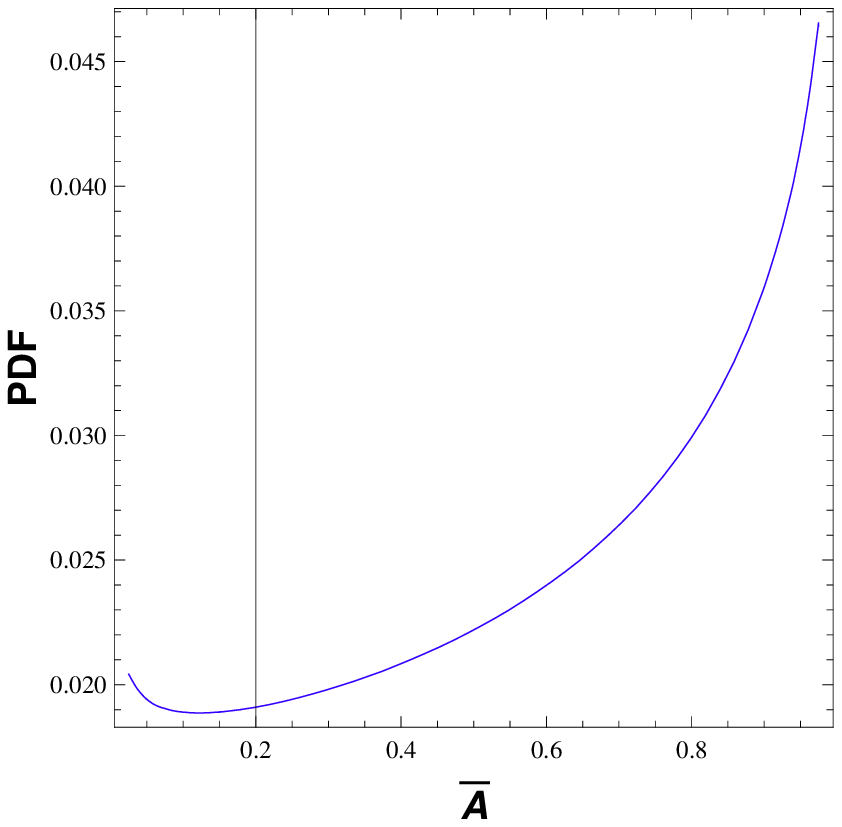}
\end{minipage} \hfill
\caption{{\protect\footnotesize One-dimensional probability distribution for $\bar A$ using only
the 2dFGRS data (left), only the SN Ia data (center) and the joint probability from both sets of data.}}
\end{figure}
\end{center}
\par
Now, we look for a self-interacting scalar field that mimics the Chaplygin gas model in presence of
a baryonic component. The equations of motion are now the following:
\begin{eqnarray}
3\biggr(\frac{\dot a}{a}\biggl)^2 &=& 8\pi G \rho + \frac{\dot\phi^2}{2} + V(\phi), \\
2\frac{\ddot a}{a} + \biggr(\frac{\dot a}{a}\biggl)^2 &=& - \frac{\dot\phi^2}{2} + V(\phi),\\
\ddot\phi + 3\frac{\dot a}{a}\dot\phi &=& - V_\phi(\phi), \quad \rho = \frac{\rho_0}{a^3}.
\end{eqnarray}
This system of equations is equivalent to the system (\ref{fe1},\ref{fe2}) if the derivative of the $\phi$ and the potential $V$ are given
in terms of the scale factor as
\begin{eqnarray}
\label{r1}
\dot\phi &=& \frac{\sqrt{3\Omega_{c0}(1 - \bar A)}}{a^3}\frac{1}{\biggr[\bar A + \frac{1 - \bar A}{a^6}\biggl]^{1/4}},\\
\label{r2}
V &=& \frac{3}{2}\Omega_{c0}\frac{2\bar A  + \frac{(1 - \bar A)}{a^6}}{\sqrt{\bar A + \frac{1 - \bar A}{a^6}}}.
\end{eqnarray}
These relations allow to define implicitly $V$ as function of $\phi$. In fact, using the Friedmann equation, it is possible to
integrate (\ref{r1}). However, apparently there is no way to have a closed expression for $V$ as function of $\phi$, except in the
case the baryonic component is absent \cite{moschella,bilic}. But, for our purposes, all we need are the relations (\ref{r1},\ref{r2}).
\par
We perturb now the Einstein-scalar field system using again the synchronous gauge condition. Following the same steps as before, and
re-expressing the final equations in terms of the variable $a$ (the scale factor) we end up with the following set of coupled equations:
\begin{eqnarray}
\delta'' + \biggr\{\frac{2}{a} + \frac{g}{f^2}\biggl\}\delta' - \frac{3}{2}\frac{\Omega_m}{f^2}\delta &=& 2\frac{l}{f}\lambda' - \frac{V_\phi}{f^2}\lambda,\\
\lambda'' + \biggr\{\frac{3}{a} + \frac{g}{f^2}\biggl\}\lambda' + \biggr\{\frac{k^2l_0^2}{a^2f^2} + \frac{V_{\phi\phi}}{f^2}\biggl\}\lambda &=&
\frac{l}{f}\delta'.
\end{eqnarray}
In these expressions, $\Omega_m$, $f$ and $g$ are defined as before, $l(a) = \dot\phi$, with $\dot\phi$ given by
(\ref{r1}) and $\lambda = \delta\phi$. Moreover, we have the following definitions:
\begin{eqnarray}
V_\phi &=& \frac{f}{l}V_a,\\
V_{\phi\phi} &=& \frac{f^2}{l^2}V_{aa} + \frac{f}{l}\biggr\{\frac{f'}{l} - \frac{l'f}{l^2}\biggl\}V_a.
\end{eqnarray}
The subscript $a$ indicates derivatives with respect to the variable $a$.
\par
This system has the advantage of admitting a straighforward extension to the generalized Chaplygin gas model without present
instability problems when $\alpha$ is negative. In fact, as in the case of any usual scalar
theory, the Laplacian term that generates
the dependence on the wavenumber $k$ does not contain the sound velocity, as it happens in the fluid reprsentation. However, at the same time, this
scalar version of the Chaplygin gas model leads to complete different predictions concerning the parameter estimations.
The figures $3$, $4$ and $5$ show the two-dimensional and one-dimensional PDF for the parameters $\Omega_{m0}$ and $\bar A$, using only the 2dFGRS or
the SN Ia data, and when the joint probability are considered. The SN Ia results are the same as before, since it is sensitive to the background
only. But, the results for the matter parameter density using the 2dFGRS are completely different: they predict a small value
for $\Omega_{m0}$, as it is obtained
for the $\Lambda$CDM or quintessence models, together with a smaller pic of probability near $\Omega_{m0} = 1$. The 2dFGRS data favors
a small, but non null, value for $\bar A$ while the SN Ia favors a value of $\bar A$ near $1$. The joint probability gives a maximum value for the probability
for values of $\bar A$ near $0.8$.
\begin{center}
\begin{figure}[!t]
\begin{minipage}[t]{0.3\linewidth}
\includegraphics[width=\linewidth]{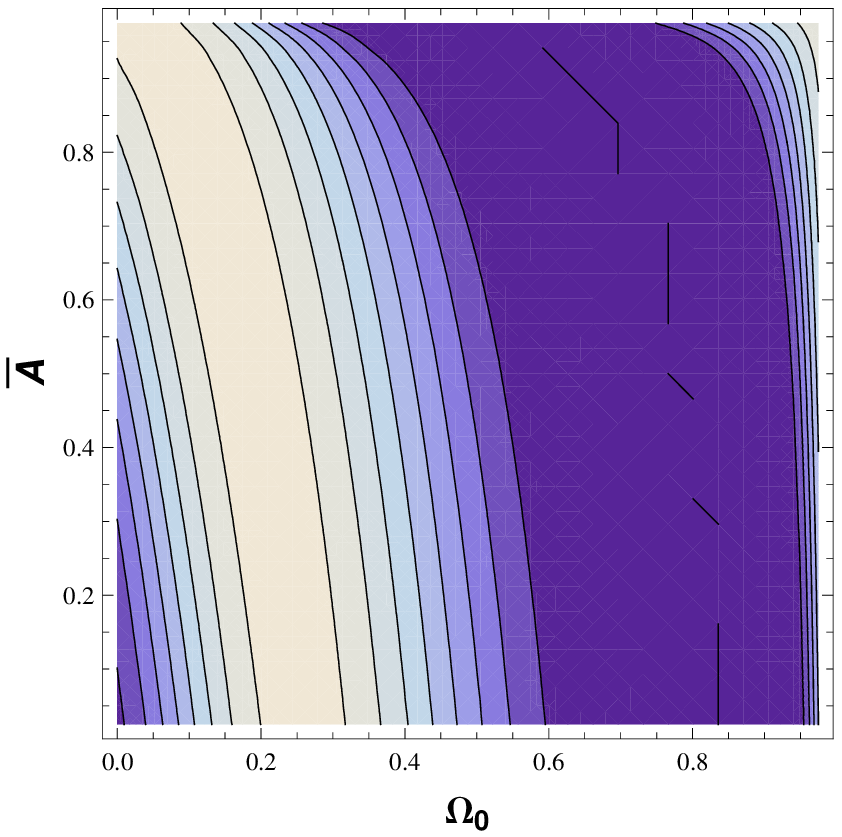}
\end{minipage} \hfill
\begin{minipage}[t]{0.3\linewidth}
\includegraphics[width=\linewidth]{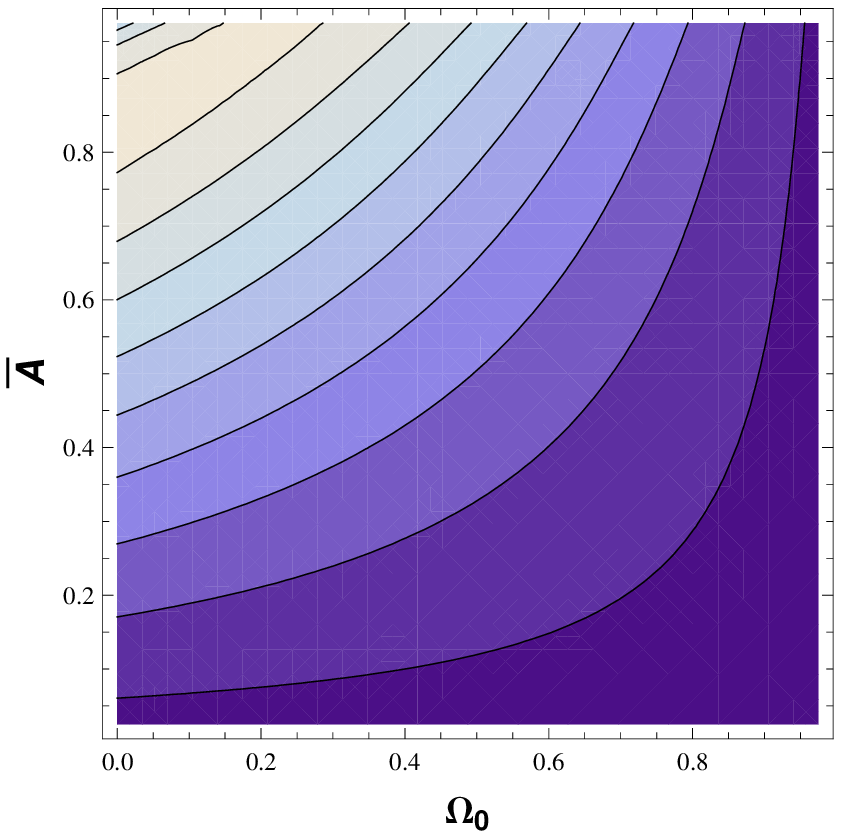}
\end{minipage} \hfill
\begin{minipage}[t]{0.3\linewidth}
\includegraphics[width=\linewidth]{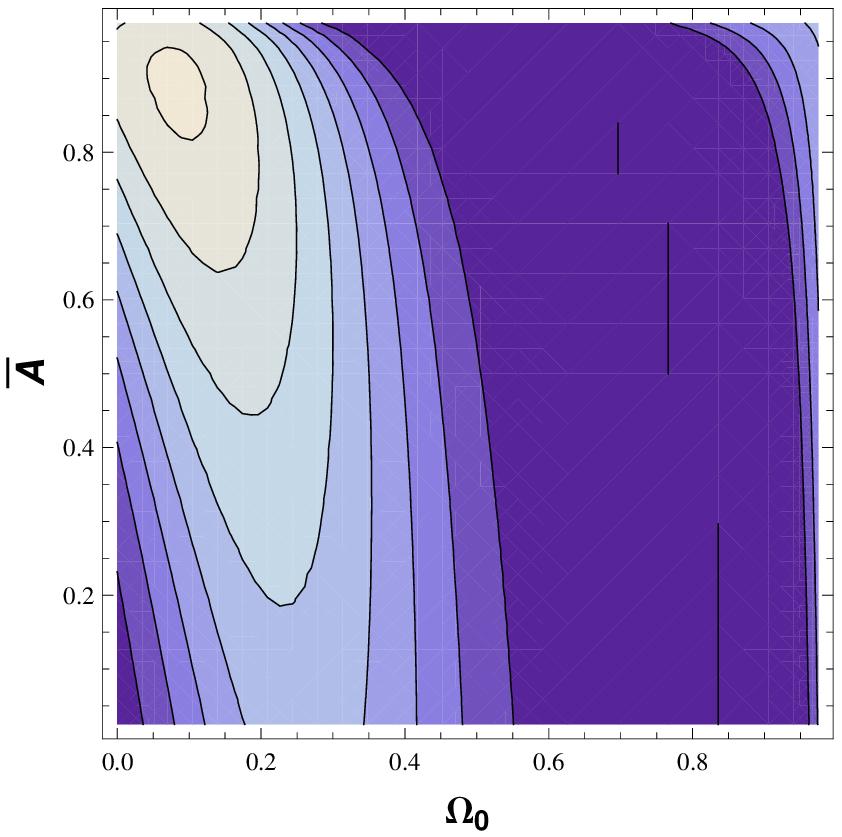}
\end{minipage} \hfill
\caption{{\protect\footnotesize Two-dimensional probability distribution for $\Omega_{0}$ and $\bar A$ using only
the 2dFGRS data (left), only the SN Ia data (center) and the joint probability from both sets of data.
The lighter is the region, the higher is
the associated probability density.}}
\end{figure}
\end{center}
\begin{center}
\begin{figure}[!t]
\begin{minipage}[t]{0.3\linewidth}
\includegraphics[width=\linewidth]{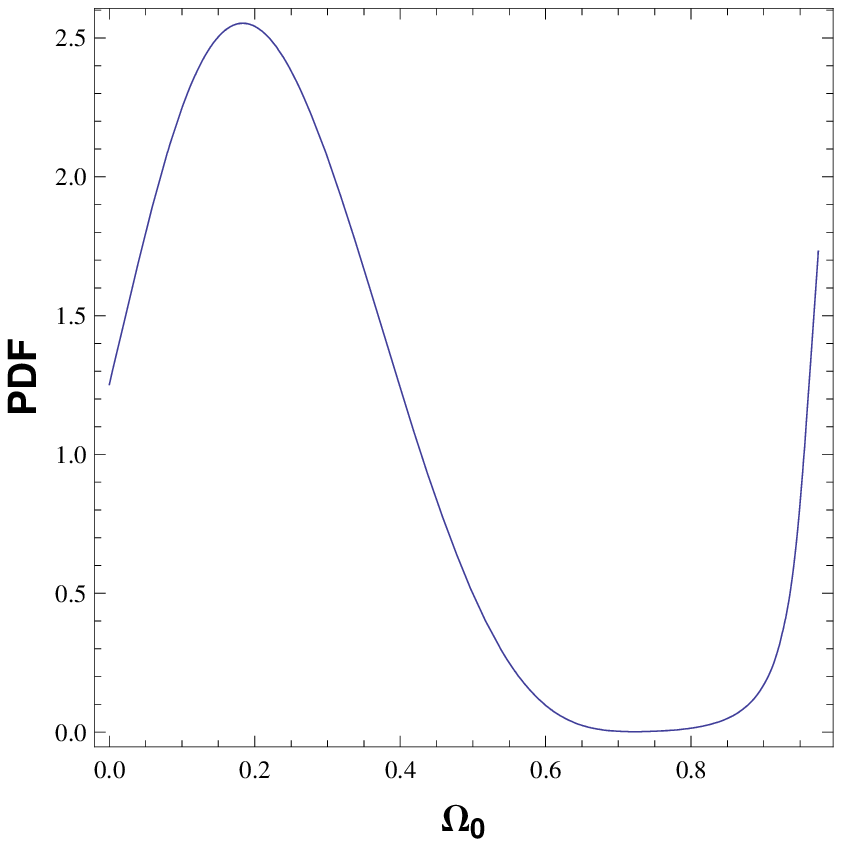}
\end{minipage} \hfill
\begin{minipage}[t]{0.3\linewidth}
\includegraphics[width=\linewidth]{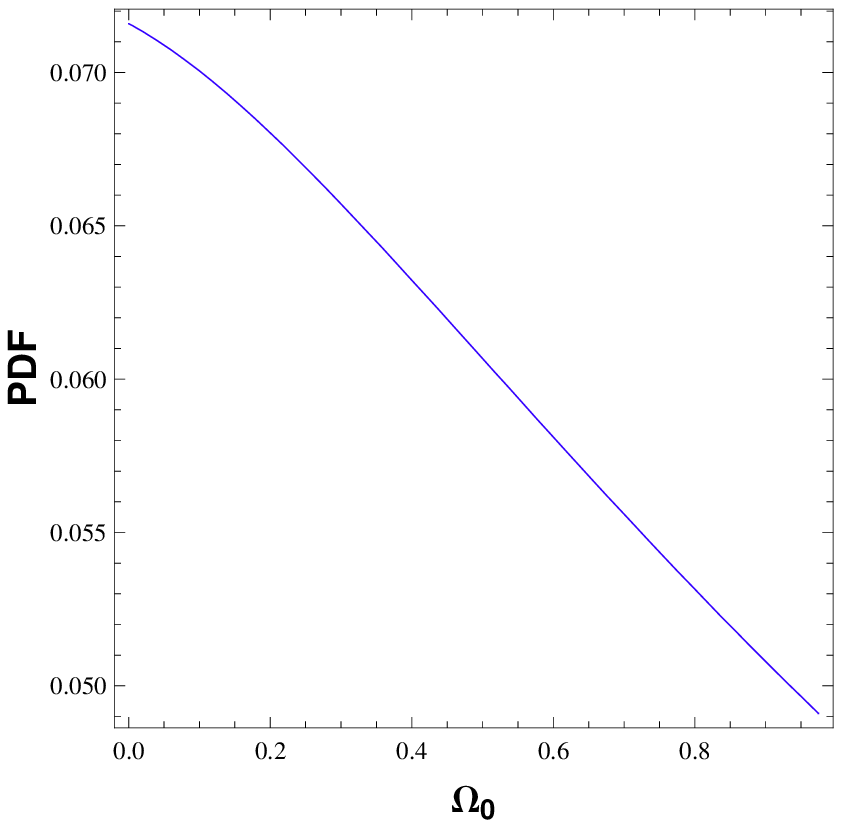}
\end{minipage} \hfill
\begin{minipage}[t]{0.3\linewidth}
\includegraphics[width=\linewidth]{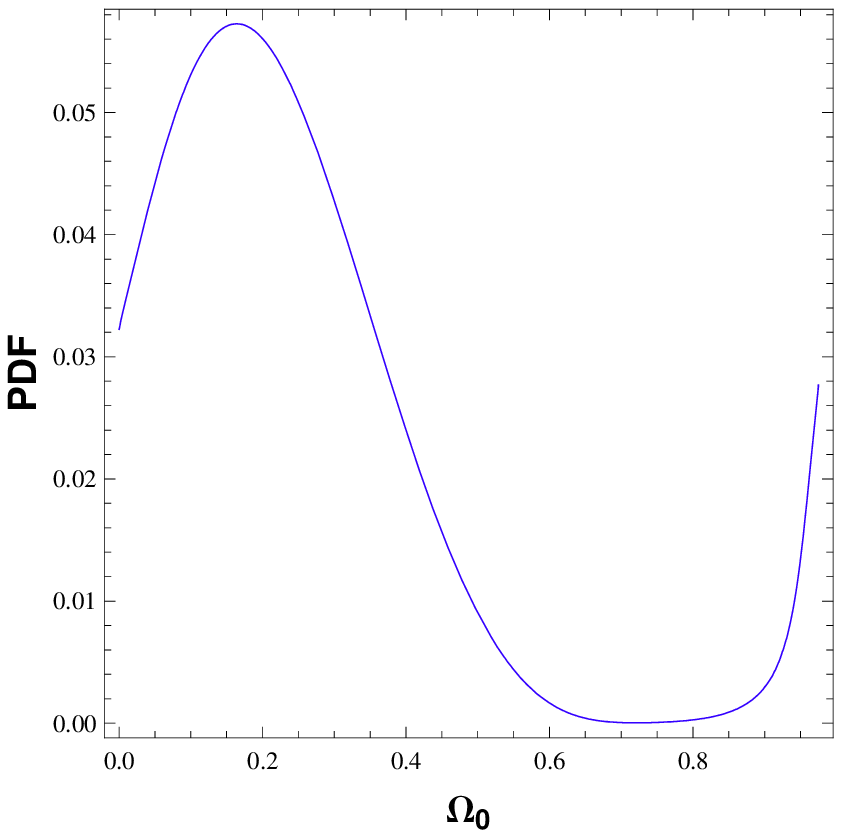}
\end{minipage} \hfill
\caption{{\protect\footnotesize One-dimensional probability distribution for $\Omega_{0}$ using only
the 2dFGRS data (left), only the SN Ia data (center) and the joint probability from both sets of data.}}
\end{figure}
\end{center}\begin{center}
\begin{figure}[!t]
\begin{minipage}[t]{0.3\linewidth}
\includegraphics[width=\linewidth]{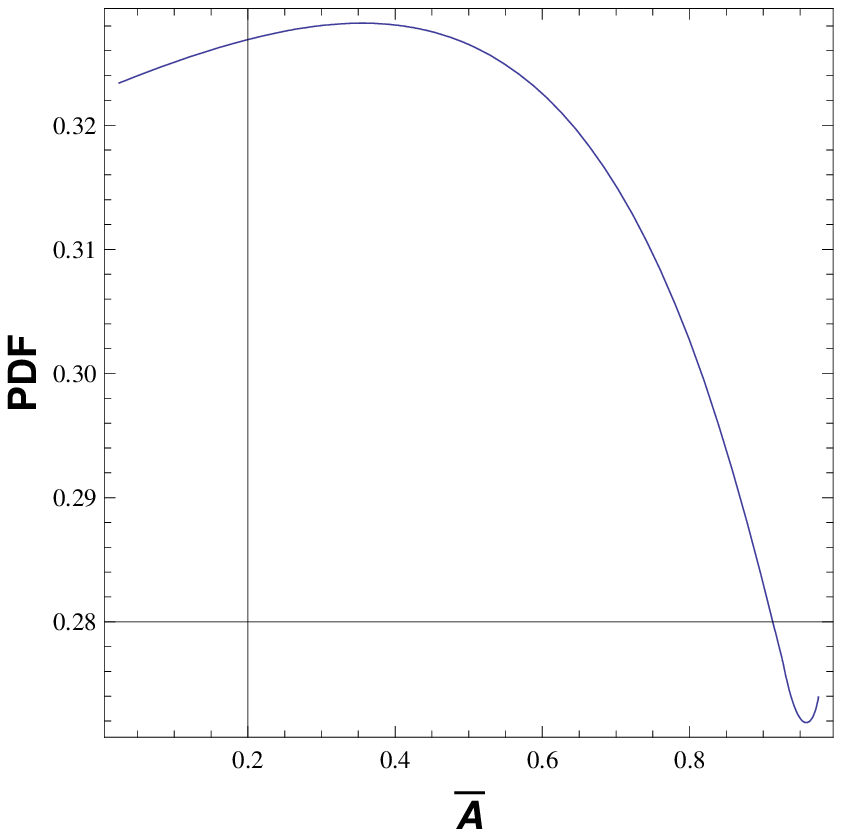}
\end{minipage} \hfill
\begin{minipage}[t]{0.3\linewidth}
\includegraphics[width=\linewidth]{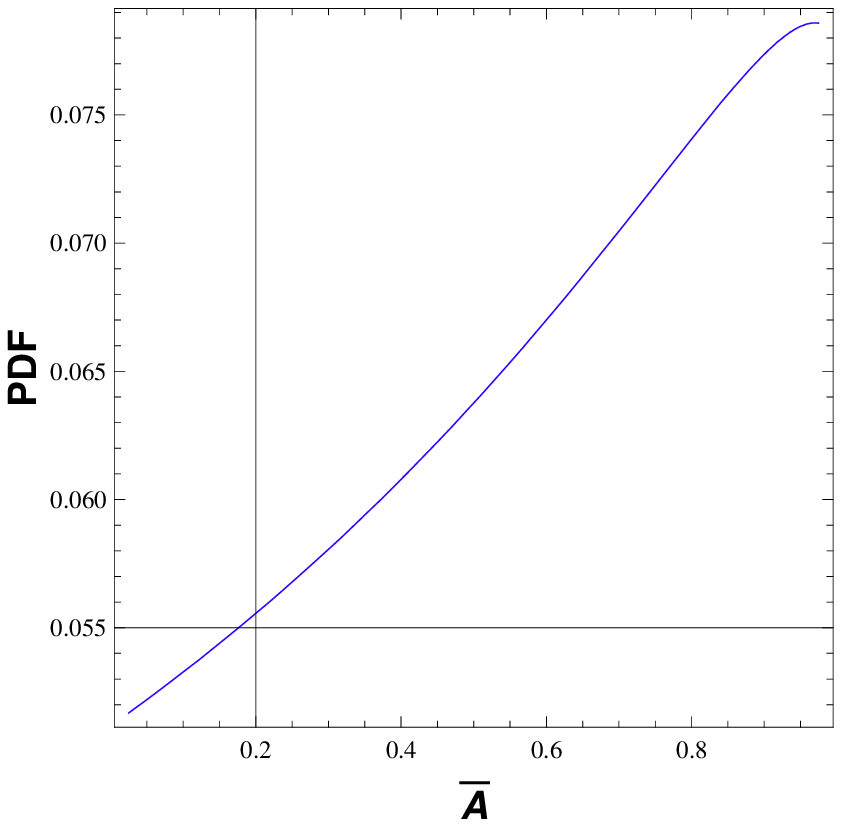}
\end{minipage} \hfill
\begin{minipage}[t]{0.3\linewidth}
\includegraphics[width=\linewidth]{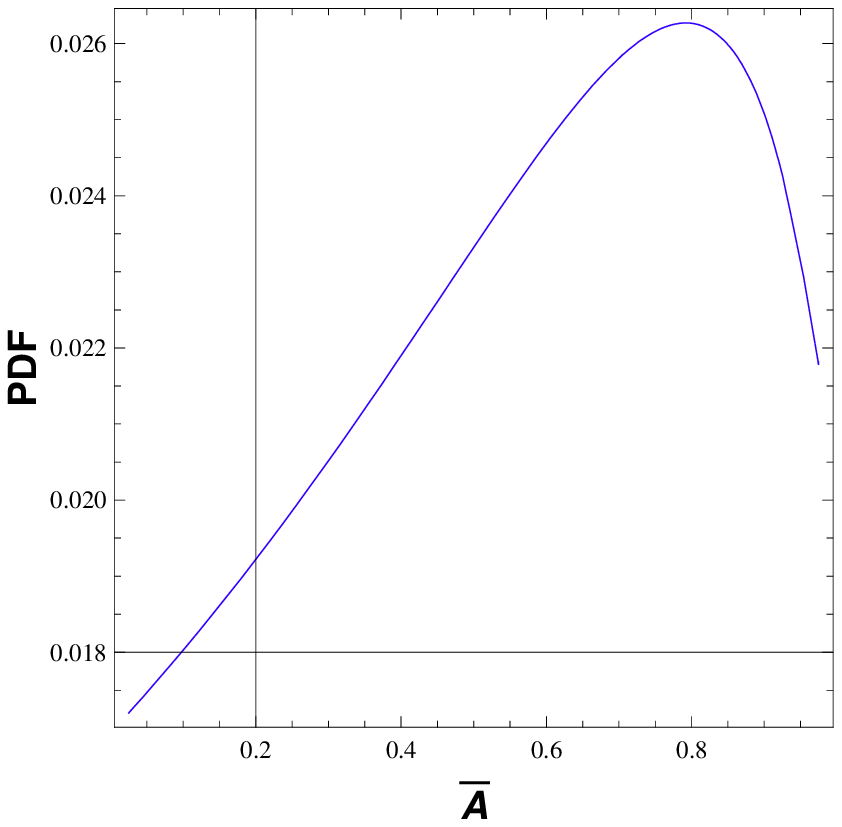}
\end{minipage} \hfill
\caption{{\protect\footnotesize One-dimensional probability distribution for $\bar A$ using only
the 2dFGRS data (left), only the SN Ia data (center) and the joint probability from both sets of data.}}
\end{figure}
\end{center}
\par
Hence, the scalar version of the Chaplygin gas model keeps the property of the original formulation, in terms of
the tachyonic field, only for the background. As a consequence it seems that in the analysis of the structure formation, and
considering the generalized Chaplygin gas model, we must consider
a space parameter where only the positive values of $\alpha$ are allowed. This may force also to restrict the SN Ia analysis
to this region of the space parameter. Of course, the final estimations for the cosmological parameter sare modified by
this restriction (see discussion in \cite{sn1}). We hope to present a full analysis of this issue in a future paper.
\newline
{\bf Acknowledgement:} J.C.F. and C.E.M.B. thanks CNPq (Brazil) and FAPES (Brazil) for partial financial support.
M.M. thanks Grant-in-Aid for 
Young Scientists (B) (No. 18740164) from Japanese Ministry 
of Education, Culture, Sports, Science and Technology for partial financial support. M.M. and J.C.F. thank also the
Institut d'Astrophysique de Paris (France) for kind hospitality during part of the elaboration of this work. We thank Winfried Zimdahl and Wiliam Ricald for their comments
and suggestions.


\begin{thebibliography}{90}
\bibitem{chaplygin} S. Chaplygin, Sci. Mem. Moscow Univ. Math. Phys. {\bf 21}, 1(1904).
\bibitem{aero} H.-S. Tsien, J. Aeron. Sci. {\bf 6}, 399(1939); T. von Karman, J. Aeron. Sci. {\bf 8}, 337(1941).
\bibitem{super} J. Hoppe, hep-th/9311059; R. Jackiw and A. P. Polychronakos, Phys. Rev. {\bf D62}, 085019 (2000).
\bibitem{jackiw} R. Jackiw, A particle field theorist�s lectures on supersymmetric, non abelian fluid mechanics and dbranes,
physics/0010042.
\bibitem{moschella} A.Y. Kamenshchik, U. Moschella and V. Pasquier, Phys. Lett. {\bf B511}, 265(2001).
\bibitem{bertolami} M.C. Bento, O. Bertolami and A.A. Sen, Phys. Rev. {\bf D66}, 043507 (2002).
\bibitem{s1} A. Riess et al, Astron. J. {\bf 116}, 1009(1998).
\bibitem{s2} S. Perlmutter et al, Astrophys. J. {\bf 517}, 565(1999) 
\bibitem{sn1} R. Colistete Jr. and J.C. Fabris, Class. Quant. Grav. {\bf 22}, 2813(2005).
\bibitem{piattella} V. Gorini, A.Y. Kamenshchik, U. Moschella, O.F. Piattella and A.A. Starobinsky, JCAP 02, 016(2008).
\bibitem{hermano} J.C. Fabris, S.V.B. Gon\c{c}alves, H.E.S. Velten and W. Zimdahl,
Phys. Rev. {\bf D78}, 103523(2008).
\bibitem{martin} J.C. Fabris and J. Martin, Phys. Rev. {\bf D55}, 5205(1997).
\bibitem{moschellabis}  Z. Keresztes, L.\`A. Gergely, V. Gorini, U. Moschella and A.Yu. Kamenshchik, Phys. Rev. {\bf D79}, 083504(2009). 
\bibitem{sola} J.C. Fabris, I.L. Shapiro and J. Sola, JCAP 02, 016(2007).
\bibitem{bilic} N. Bilic, G.B. Tupper and R.D. Viollier, Phys. Lett. {\bf B535}, 17(2002).
\end{thebibliography}
\end{document}